\documentclass[12pt]{article}

\usepackage{newtxtext,newtxmath}

\usepackage{graphicx}

\usepackage{soul}

\usepackage[letterpaper,margin=1in]{geometry}

\renewenvironment{abstract}
	{\quotation}
	{\endquotation}

\date{}

\makeatletter
\renewcommand{\fnum@figure}{\textbf{Figure \thefigure}}
\renewcommand{\fnum@table}{\textbf{Table \thetable}}
\makeatother

\usepackage{scicite}

\usepackage{url}

\usepackage{xcolor}
\usepackage{soul}

\newcommand{\nmss}{nm\,s$^{-2}$}
\newcommand{\kgmmm}{kg\,m$^{-3}$}

\def\scititle{
	{Unblockable Communication With Gravity}
    }
\title{\bfseries \boldmath \scititle}

\author{
	Andrew~J.~Groszek$^{1\ast\dagger}$,
	Charles~W.~Woffinden$^{1\dagger}$,
    Michael~D.~Harvey$^{1}$,
    \and
	Andrew~G.~White$^{1}$,
    Matthew~J.~Davis$^{1\ast}$
    \and
	\small$^{1}$Australian Research Council Centre of Excellence for Engineered Quantum Systems,
    \and 
    \small School of Mathematics and Physics,
    University of Queensland, St. Lucia, QLD 4072, Australia.
    \and
	\small$^\ast$Corresponding authors. Email: a.groszek@uq.edu.au; mdavis@uq.edu.au\and
	\small$^\dagger$These authors contributed equally to this work.
    }

\begin{document} 

\maketitle

\begin{center}
    A scheme for wireless information transfer based on mass motion and gravitational‑field detection is proposed and demonstrated.
\end{center}

\begin{abstract} \bfseries \boldmath
All modern wireless communication technologies are based on electromagnetism. However, electromagnetic signals are readily susceptible to screening and blocking, so their availability cannot be guaranteed in adverse environments.
This raises a fundamental question: Can information be transmitted through an unblockable channel? Here we show that gravity, under physically realistic conditions, offers such a path. We propose and implement a wireless communication protocol in which a broadcaster encodes a binary message by moving a mass, while a receiver detects the resulting gravitational signal with a gravimeter. We validate this scheme experimentally, successfully transmitting a gravitational message a distance of $\approx$~$0.7$\,m through a brick wall at a rate of 1\,bit\,min$^{-1}$. These results establish gravity as a viable platform for practically unblockable communication.
\end{abstract}

\noindent

\section*{Introduction}

Suppose Alice is confined to a perfectly shielded room: no photons, sound waves, particles, or thermal signals can enter or escape. If Bob is outside the room, is there any physical mechanism Alice can use to communicate with him? The answer is yes---she can use gravity. Since all mass is non-negative~\cite{schoen_proof_1979, de_aguiar_alves_positive_2025}, there is no gravitational analogue of charge cancellation and hence no way to hide or screen the presence of mass. Moreover, no known material can passively shield or attenuate gravitational fields~\cite{unnikrishnan_new_2000}. Thus, if Alice moves a mass within the room, the local gravitational field at Bob's location will change, and this effect can be detected using a gravimeter. Alice can therefore communicate with Bob by encoding information in the gravitational field. In the simplest scenario, Alice can generate a binary gravitational signal by moving the mass between two positions (Fig.~\ref{fig:alice_bob}).

This practically unblockable wireless communication scheme stands in contrast to conventional methods based on electromagnetic waves, as these can be passively blocked by Faraday cages, and attenuate when passing through matter~\cite{griffiths_introduction_2023}. Near-unblockable communication has previously been proposed using neutrinos~\cite{stancil_demonstration_2012} and gravitational waves~\cite{wang_gravitational_2026}, but these approaches are somewhat hindered by the difficulty in generating a measurable signal---e.g.~requiring the manipulation of black holes---as well as the prohibitively large detectors required~\cite{stancil_demonstration_2012, ligo_scientific_collaboration_and_virgo_collaboration_observation_2016}. By contrast, the protocol considered here does not rely on the generation or detection of propagating gravitational waves, but instead on quasi-static changes in the Newtonian gravitational field produced by controlled motion of a nearby mass. The implementation is therefore straightforward, requiring only a mass and an off-the-shelf, hand-held gravimeter. For example, if Alice has access to a $100$\,kg mass, and moves it from $1$\,m to $2$\,m away from Bob, the expected change in gravitational acceleration at Bob's location is $\sim$~$5$\,\nmss\:[see Eq.~\eqref{eq:newton_pointmass}]. 
While this change may go unnoticed by the casual observer, Bob's gravimeter is capable of detecting it. 

Off-the-shelf gravimeters, typically based on mass--spring~\cite{Goodkind_1999, VanCamp_2017, Halder_2018} or falling-corner-cube~\cite{Niebauer_2015} designs, 
can provide sensitivities of $\sim$~$1$\,\nmss~\cite{Freeden_2021}. Higher precisions of $\sim$~$0.01$\,\nmss\:can be reached with superconducting gravimeters~\cite{Virtanen_2006, Freeden_2021}, although at a cost of portability and price. Emerging gravimetry platforms---such as cold-atom interferometry~\cite{Kasevich_1992, Xuejian_2019}, optically levitated nanospheres~\cite{Monteiro_2020, liang_yoctonewton_2023}, and microelectromechanical systems devices~\cite{Middlemiss_2016, Tang_2019, Mustafazade2020}---promise even higher precision or low-cost alternatives in the near future. 
These advances may unlock a range of previously inaccessible technologies, such as detecting contraband materials~\cite{kirkendall_imaging_2007} and underwater objects~\cite{Wu_2010}, or constructing gravimetry networks for early-warning earthquake detection~\cite{Kwon_2020}. 
It is therefore an opportune time to consider the possibility of gravitational communication.

Here, we first explore the theoretical properties of gravitational communication,
including signal strengths, broadcast directionality, data rates, and robustness. We then experimentally demonstrate this scheme by broadcasting a gravitational message using an elevator counterweight, with an off-the-shelf spring-based gravimeter used for detection.

\section*{Results}

\subsection*{Signal generation}
Our proposed gravitational communication scheme is purely classical, and can therefore be described by Newton's law of gravitation. 
If Alice holds a point mass $M$ at position $\mathbf{r}$, the gravitational acceleration it creates at Bob's detector (assumed to be at the origin) is
\begin{equation} \label{eq:newton_pointmass}
    \mathbf{g}(\mathbf{r}) = \frac{GM}{r^2}\hat{\mathbf{r}},
\end{equation}
where $G$ is the gravitational constant. As Alice moves her mass between two positions, $\mathbf{r}_0$ and $\mathbf{r}_1$, this gravitational acceleration changes by $\Delta \mathbf{g} = \mathbf{g}(\mathbf{r}_1) - \mathbf{g} (\mathbf{r}_0)$, generating a signal for Bob to measure.

According to Eq.~\eqref{eq:newton_pointmass}, the largest signal that Alice can theoretically generate is $|\Delta \mathbf{g}| = GM/d^2$, assuming she is constrained to one side of the sensor (as in Fig.~\ref{fig:alice_bob}), and that the room is large enough for her to move the mass from some initial distance $r_0 = d$ to a much greater $r_1 \gg d$. 
However, if Bob's gravimeter---like many off-the-shelf gravimeters---only measures the vertical component of gravity $g_z$, the maximum signal will decrease. In the best case, Alice can still achieve $|\Delta g_z| \approx 0.8 GM/d^2$, provided she is able to move the mass between vertical positions $z_\pm^* \approx \pm 0.7 d$ (details of this calculation can be found in the Supplementary Materials). Geometric effects can therefore be largely mitigated, but only by displacing the mass over a distance on the order of the transmission length. 

In the above we have implicitly assumed that (i) Alice and Bob share a fixed inertial reference frame, and (ii) Alice’s mass can be moved without appreciably affecting the reference frame. This is an excellent approximation when the frame is anchored to a planet-sized body such as the Earth, for which the recoil associated with Alice’s mass displacement is negligible. In a freely floating system such as a spacecraft, however, momentum conservation requires any displacement of Alice’s mass to be accompanied by compensating motion elsewhere in the system, so that the centre of mass remains fixed. This compensation could in principle be supplied by thrusters, which eject a small amount of propellant carrying the required momentum. We leave a full treatment of gravitational communication in spacecraft for future work.

\subsection*{Directional properties}
By moving the mass between her two chosen positions, Alice has created a gravitational antenna. To understand this antenna's properties, we consider a point mass moved vertically between $+\Delta z$ and $-\Delta z$, as illustrated in Fig.~\ref{fig:antenna_properties}A, and a single $z$-axis gravimeter as the detector, located at polar position $\mathbf{r} = (r, \alpha)$.  Figure~\ref{fig:antenna_properties}B shows a series of normalised polar plots of the change in gravitational signal as a function of the gravimeter's angular position $\alpha$ for multiple radii $r$ (denoted by the legend). For mathematical details, see Supplementary Materials.

For $r / \Delta z \ll 1$, the signal is independent of $\alpha$, and hence the antenna is omnidirectional for an observer near its centre. As $r / \Delta z$ approaches unity, the signal becomes increasingly directional along the $z$-axis, due to the increasing proximity of the gravimeter to the mass at $\alpha = 0^\circ$ and $\alpha=180^\circ$ (note that the signal diverges at $r / \Delta z = 1$ when the two coincide). Further increasing $r$ leads to the far-field regime ($r/\Delta z \gg 1$), where the antenna's transmission broadens again, but remains primarily directed along its axis.

For comparison, we show the far-field radiation pattern of an electromagnetic (EM) half-wave dipole antenna~\cite{ramo1994} (dotted line Fig.~\ref{fig:antenna_properties}B). This looks broadly similar to the gravitational antenna's far-field signal, except for a rotation by $90^\circ$ relative to the antenna's axis. However, the gravitational antenna transmits at $0.25$ of its maximum perpendicular to its two lobes in the far-field, whereas the dipole EM antenna has zero transmission in the perpendicular directions. Despite the similarities between the radiation patterns, we emphasise that the two physical scenarios are quite different: EM antennas rely on flowing currents to generate EM waves, whereas the quasi-static gravitational antenna creates a change in a static field, more akin to a static electric dipole. Accordingly, the gravitational antenna signal decays with distance as $\sim r^{-3}$ in the far field (see Supplementary Materials), compared to the $\sim r^{-1}$ decay of an EM antenna~\cite{griffiths_introduction_2023}.

\subsection*{Data rate}

A key property of any communication channel is the maximum rate of error-correctable data transfer it allows, known as its capacity $C$. For a noisy digital channel, the Shannon--Hartley theorem~\cite{Shannon_1948} gives
$C = (f_\mathrm{t}/2) \log_2 ( 1 + \mathrm{SNR})$, with $f_\mathrm{t}$ the bit transmission frequency and SNR the signal-to-noise ratio. Here we take $f_\mathrm{t} = 1/(T + \tau)$, where $T$ is the time it takes Alice to move the mass between positions 0 and 1, and $\tau$ is the integration time of Bob's measurement. Assuming the noise is Gaussian, the SNR will grow as $\approx (f_\mathrm{s} \tau)^{1/2}$ for sampling frequency $f_\mathrm{s}$. For a gravitational signal from a point mass, the SNR will also decay with distance $r$ as $\mathrm{SNR}_\mathrm{ss} (r/r_\circ)^{-2}$, where $\mathrm{SNR}_\mathrm{ss}$ is the single-shot SNR at some reference distance $r_\circ$. Putting this all together, the channel capacity for binary gravitational communication can be written:
\begin{equation} \label{eq:shannon2}
    C = \frac{1}{2(T + \tau)} \log_2 \left[ 1 + \mathrm{SNR}_\mathrm{ss} \frac{\sqrt{f_\mathrm{s} \tau}}{(r/r_\circ)^2} \right].
\end{equation}
In the regime where the signal is easily resolvable (i.e.~$\mathrm{SNR} \gg 1$), the logarithmic term varies slowly with both integration time $\tau$ and distance $r$, and hence the prefactor dominates the scaling. In this limit, reducing $r$ therefore makes little difference to the channel capacity---it is ultimately limited by $T + \tau$, and would be most efficiently increased by reducing these two timescales as much as possible. 
Conversely, in the limit of near-zero signal, $\mathrm{SNR} \gtrsim 0$, the channel capacity reduces to $C \sim \tau^{1/2}/(T + \tau)$. This function is maximised at $\tau = T$, so reducing $\tau$ below this value actually leads to a reduction in the channel capacity, despite increasing the bit rate.

\subsection*{Robustness to signal jamming, blocking and spoofing}

While this communication method cannot be passively blocked in the manner of electromagnetic shielding, it can in principle be affected by active signal degradation strategies. A malicious actor, Mallory, could attempt to jam the signal by moving other nearby masses or generating environmental vibrations, reducing the signal-to-noise ratio at Bob's gravimeter. Mallory could also attempt to locally block (spoof) Alice’s signal by moving a nearby mass so as to counteract (reproduce) the change in gravitational field $\Delta \mathbf{g}$ at Bob’s location. In this case, Bob would observe a modified or misleading signal rather than the one generated by Alice.

In a more elaborate scenario, Mallory could attempt to block the signal entirely by constructing the gravitational analogue of a Faraday cage. This would require surrounding either Alice or Bob with a continuous distribution of controllable mass and dynamically reconfiguring it in response to Alice’s motion so as to cancel $\Delta \mathbf{g}$ at all times. Such an arrangement, while not forbidden by physics, would require Mallory to have extraordinary capabilities with regard to control and coordination of mass distributions. We explore this possibility further in the Supplementary Materials.

\subsection*{Experiment}
We experimentally demonstrate the proposed method of communication by transmitting a gravitational message through a wall from an elevator shaft into an adjacent room using a movable elevator counterweight. The change in gravitational field is detected at a nearby location with an off-the-shelf sensor. Specifically, we use a LaCoste and Romberg Model D mass--spring gravimeter, which measures relative changes in the vertical component of the gravitational field~\cite{landr}. It has a data acquisition rate of $\approx 1.9$\,Hz, digital resolution of $10$\,\nmss, and an in-built low-pass filter with a $40$\,s window for removing high frequency noise. We have measured its sensitivity in the experimental location to be $\sim 2$\,\nmss\:after $30$\,s of averaging (see Supplementary Materials).

\begin{figure}[b]
    \centering
    \includegraphics[width=0.5\columnwidth]{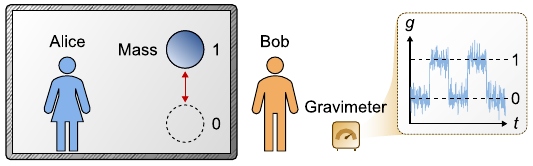}
    \caption{\textbf{Schematic of gravitational communication.} Alice (inside a perfectly shielded room) can communicate with Bob (outside the room) by moving a mass between two positions 0 and 1 to create a binary change in the local gravitational field. The inset shows a schematic of the gravitational acceleration $g$ measured with Bob's gravimeter as Alice moves the mass between the two positions.}
    \label{fig:alice_bob}
\end{figure}

\begin{figure}[b]
    \centering
    \includegraphics[width=0.5\columnwidth]{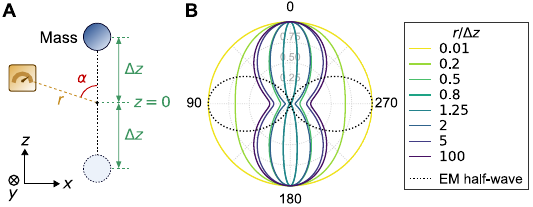}
    \caption{\textbf{Gravitational dipole antenna.} (\textbf{A}) Schematic of a point mass moving between vertical positions $+\Delta z$ and $-\Delta z$. The gravimeter, at position $(r,\alpha)$ in the $xz$-plane, measures a change in gravitational field as the mass moves between the two positions. (\textbf{B}) Polar plots of the vertical signal $\Delta g_z$ from the gravitational antenna, as a function of the polar angle $\alpha$ from the axis of the antenna. Line colours denote the radial distance $r/\Delta z$ from the centre of the antenna, indicated by the legend. The strength of the signal at each radius is normalised to the largest value at that radius. For comparison, an electromagnetic half-wave dipole antenna radiation pattern is shown as a black dotted line. Note that all patterns are symmetric with respect to rotation around the $z$-axis.}
    \label{fig:antenna_properties}
\end{figure}

\begin{figure}[b]
    \centering
    \includegraphics[width=0.5\columnwidth]{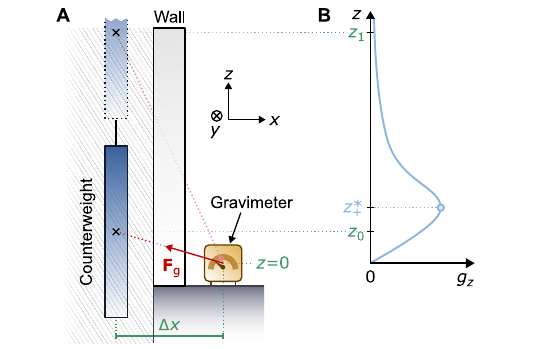}
    \caption{\textbf{Schematic of experimental setup}. (\textbf{A}) The counterweight in the lower (0) and upper (1; dotted boundary) position (not to scale). The force $\mathbf{F}_\mathrm{g}$ on the gravimeter's test mass due to the gravitational attraction to the counterweight changes in both magnitude and direction between the two positions, as indicated. (\textbf{B}) Schematic of the expected vertical gravitational signal $g_z$ as a function of the counterweight's vertical position. The lower and upper counterweight positions are indicated as $z_0$ and $z_1$, respectively. The maximal vertical signal occurs at an intermediate height $z_+^*$, resulting from the tradeoff between the directionality of the sensor and the $\sim z^{-2}$ fall-off of gravity.}
    \label{fig:schematic}
\end{figure}

We placed the sensor on the ground floor, separated from the counterweight by a brick wall that blocks acoustic, optical and near-infrared communication (see Fig.~\ref{fig:schematic}A). As the counterweight moves vertically, there is a change in the vertical gravitational acceleration $g_z$ at the gravimeter's location. This change, schematically depicted in Fig.~\ref{fig:schematic}B as a function of the counterweight's vertical position $z$, allowed us to encode a binary message in the movement of the mass as it transited between the lower and upper floors (as in Fig.~\ref{fig:alice_bob}).

To maximise the signal for the demonstration, we placed the gravimeter as close as possible to the counterweight, at horizontal distances $\Delta x = 0.56(3)$\,m and $\Delta y = 0.00(5)$\,m from the gravimeter's centre to the counterweight centre of mass (COM). We estimated the counterweight's COM to be $z_0=0.39(6)$\,m above the gravimeter in the lower (0) position, and $z_1=5.80(6)$\,m above it in the upper (1) position (see Fig.~\ref{fig:schematic}). The COM of the counterweight was therefore a distance of $0.68(4)$\,m away from the sensor in its closest position, which we define as the range of the communication scheme. We also estimated the counterweight's mass to be $1800(200)$\,kg. Materials and Methods contain further details regarding these measurements.

We constructed a model of this setup, approximating the counterweight as a rectangular prism of uniform density. Using an integral form of Newton's law of gravitation, we calculated the vertical gravitational signal $g_z$ at the location of the sensor as a function of the model counterweight's vertical position (see schematic in Fig.~\ref{fig:schematic}B; further details can be found in Materials and Methods). The model predicts a change in $g_z$ of $50(15)$\,\nmss\:as the counterweight moves between the upper and lower positions, which we expect to be resolvable with our gravimeter. The model further suggests that a $\approx 70\%$ signal boost would be achievable by optimising the counterweight position such that $z_0 = z_+^*$ (see Fig.~\ref{fig:schematic}B).

We began the experiment by vertically aligning the gravimeter using its built-in spirit levels, followed by $\approx 20$ minutes of settling time. We then started acquiring data, leaving the elevator counterweight in the lower position for $\approx 25$ minutes to obtain a background measurement. Following this, we transmitted a gravitational sequence of nine $7$-bit ASCII characters by moving the elevator counterweight between its two positions in a chosen sequence. The bits were sent at a rate of one per minute, allowing for the $\approx 25$\,s counterweight movement time, followed by $10$\,s settling time, and $25$\,s of integration time. We chose to transmit the word `Gravity', which is expressed as 49 bits: 1000111 1110010 1100001 1110110 1101001 1110100 1111001. We also included a control sequence of seven alternating bits (1010101) both before and after the message to delineate the signal from the background and establish the transmission bit rate the receiver should expect. After the bit sequence, we acquired an additional $\approx 20$ minutes of background (with the counterweight returned to its lower position) in order to quantify measurement drift (see Supplementary Materials).

\begin{figure}[t]
    \centering
    \includegraphics[width=1\linewidth]{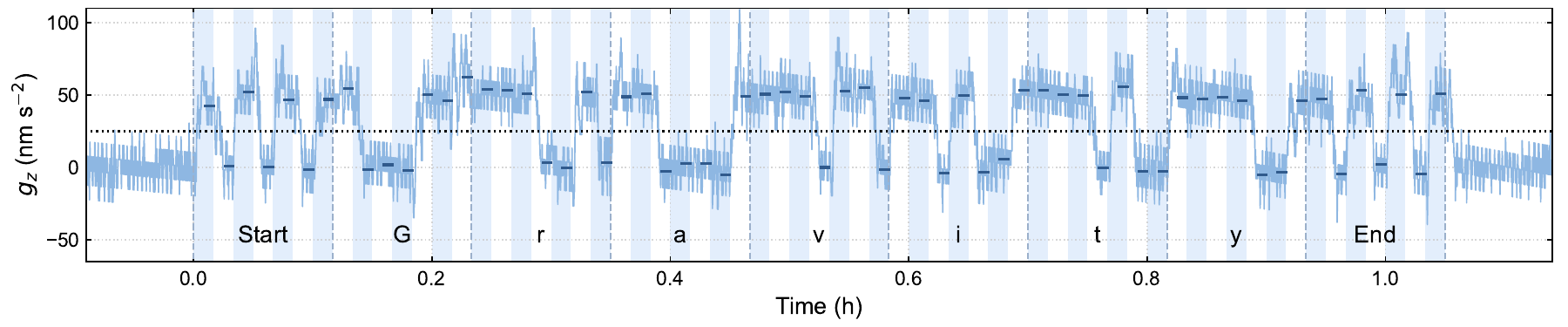}
    \caption{\textbf{Experimental gravitational signal.} Gravimetry signal $g_z$ over the transmission sequence (corrected for tide and drift; see Supplementary Materials). The start and end points of each 7-bit ASCII character are denoted with vertical dashed lines. The $60$\,s temporal window for each bit is indicated by the alternating blue and white shading, and horizontal solid lines show the corresponding mean $g_z$ for that interval, where the average is taken over the final $25$\,s. All measurements above (below) $g_z=25$\,\nmss\:(horizontal black dotted line) after correction are recorded as a $1$ ($0$). Note that this is a relative gravitational measurement, and hence the offsets of the data are arbitrary. The sign convention of the data is such that a positive change in $g_z$ corresponds to an increase in downward acceleration.}
    \label{fig:grav_data}
\end{figure}

The gravitational signal $g_z$ detected by the gravimeter over the course of the experiment is shown in Fig.~\ref{fig:grav_data}. Note that these data have been corrected for both Earth tide and sensor drift (the raw data are presented in fig.~\ref{fig:grav_data_raw}). Each character is delineated with vertical dashed lines, and each bit with alternating background shading. The horizontal solid lines indicate the mean $g_z$ value measured over that time window. Evidently, the data show distinct shifts of $\approx 50$\,\nmss\:each time the elevator counterweight is moved, consistent with the prediction of our model. While there are small fluctuations in the mean values, the averaged 0 and 1 signals are clearly distinguishable, and the detected bit sequence exactly matches the original transmission. For comparison, a dotted horizontal line is shown in Fig.~\ref{fig:grav_data} at $25$\,\nmss, approximately halfway between the two bit values. Even without averaging, the bit error rate of individual measurements (i.e.~the probability of a bit flipping by crossing the threshold) is already $\approx$~$0.01$. After averaging, this value reduces to $\approx$~$10^{-9}$ (assuming Gaussian fluctuations), comparable to optical fibre error rates~\cite{Senior_2009}. To further validate this result, we have confirmed that the measurement is insensitive to local magnetic fields, which could in principle affect the gravimeter's steel spring (see Supplementary Materials). We have thus successfully transmitted and received a message via the gravitational field.

\subsection*{Potential improvements}

Our experiment achieved a data transfer rate of $1$\,bit\,min$^{-1}$. The channel capacity for our experimental parameters is $\approx 2.4$\,bit\,min$^{-1}$, so in principle improvements could be made to the scheme by implementing more efficient encoding methods, e.g.~phase shift keying~\cite{Proakis2007}. This optimal information rate is limited by a number of factors, such as the slow movement of the elevator, and the $40$\,s low-pass filter integrated within the gravimeter. If our building had a modern elevator with a transition time of $5$\,s, and we had access to the latest Scintrex CG-6 gravimeter~\cite{cg6}, the channel capacity would increase to $\approx 15$\,bit\,min$^{-1}$ [assuming an integration time of $0.1$\,s for a single sample, with $\mathrm{SNR} \approx 5$~\cite{francis_performance_2021}]. However, improvements beyond this would be challenging with our scheme, given that the mass movement time is the limiting factor.

While the data rate remains modest even after foreseeable improvements, the communication range can in principle be increased substantially. The achievable range is set by the magnitude of the field perturbation produced at the receiver relative to the sensitivity and noise floor of the detector. Table~\ref{tab:range_optimisation} lists a number of optimisations that could be made, as well as their expected improvement factor for the range, based on Eq.~\eqref{eq:newton_pointmass}. Firstly, the configuration could be modified by compressing the mass into a more spatially localised geometry (ideally, a sphere), and vertically aligning it with the gravimeter. These changes would increase the signal by a factor of $\approx 4.9$ at the same distance of $\approx 0.7$\,m. The mass itself could be increased within reason to, e.g.~$7.5 \times 10^4$\,kg [equivalent to a typical military tank~\cite{Army_2025}]. An order of magnitude of SNR could also be sacrificed in favour of range, and in principle a higher sensitivity gravimeter could be used [e.g.~a superconducting device~\cite{Freeden_2021,Virtanen_2006}, or potentially a future cold-atom interferometer~\cite{chiow_102hbark_2011, HOSSEINIARANI_2024}]. Our experimental model predicts that sacrificing SNR (the only option available with our current setup) could increase the potential range to $\approx 1.7$\,m. Combining all the improvements listed in Table~\ref{tab:range_optimisation}, we predict that the range could be increased by a factor of $\sim 700$ to $\sim 500$\,m. 

Clearly, even with these idealised improvements, the scheme has a low data rate and short range when compared to other use cases. For example, the transmission rate for uplink commands to the Voyager~1 spacecraft---the furthest human-made object from Earth at $\sim 10^{13}$\,m away---is an appreciably higher $\approx 960$\,bit\,min$^{-1}$~\cite{Nasa_2024} (although being electromagnetic, this signal is inherently susceptible to jamming and blocking). We therefore expect gravitational communication to be best suited to the transmission of concise messages over relatively short distances in challenging or constrained environments.

\section*{Discussion}
We have proposed a form of practically unblockable communication using classical gravitation, and performed a proof-of-principle experimental demonstration. Using an elevator counterweight and an off-the-shelf gravimeter, we have successfully transmitted a gravitational message. 

Although this technology is restricted to relatively short-range, low bit-rate applications, its unique advantages may be relevant for some niche use cases. Since it is possible to measure gravitational fields without the need for electronics~\cite{kater1818, landr} (akin to semaphore), this approach could be made radiation-proof; however, unlike semaphore, it is not limited by line of sight.

A scheme similar to that demonstrated here would be well-suited to communication from within secure facilities in a covert manner, which could be considered to prevent data leaks. There may also be potential applications enabling wireless communication to or from sealed nuclear bunkers. Beyond its potential applications, this work demonstrates that gravity itself can serve as a carrier of information, albeit with limited bandwidth and range. Should the reader ever experience prolonged entrapment in a soundproofed, Faraday-caged elevator with a malfunctioning phone, the present method offers an experimentally verified means of signalling for aid.

\section*{Materials and Methods}

\subsection*{Analytical model} \label{sec:model}

We have constructed a model of the experimental configuration to estimate the gravitational signal we expect to observe at the location of the gravimeter. The counterweight is approximated to be a 3D rectangular prism of uniform of density $\rho=7300(200)$\,\kgmmm\:and volume $L_x \times L_y \times L_z$, with $L_x = 0.205(5)$\,m, $L_y = 0.82(1)$\,m, and $L_z = 1.47(4)$\,m. We position its COM at $x=0.56(3)$\,m, $y=0.00(5)$\,m, and allow it to move along the vertical ($z$) axis (see the next section for details on how these measurements have been obtained). At each vertical position, we calculate the gravitational field by integrating Newton's law of gravitation over the density distribution:
\begin{equation} \label{eq:newton_distribution}
    \mathbf{g}(\mathbf{r}_\mathrm{s}) = - G \int_\mathcal{V} \mathrm{d}^3\mathbf{r}' \frac{\rho(\mathbf{r}')}{|\mathbf{r}_\mathrm{s} - \mathbf{r}'|^3} (\mathbf{r}_\mathrm{s} - \mathbf{r}'),
\end{equation}
where $\mathcal{V}$ is the region containing nonzero density, and $\mathbf{r}_\mathrm{s}$ is the location of the sensor. For a rectangular prism of fixed density $\rho$ contained within the region $x_1 \leq x \leq x_2$, $y_1 \leq y \leq y_2$, $z_1 \leq z \leq z_2$, the resulting expression for $\mathbf{g}$ can be written analytically as~\cite{li_three-dimensional_1998}:
\begin{equation}
    \mathbf{g}(\mathbf{r}_\mathrm{s}) = -G \rho \sum_{k,l,m = 1}^{2} (-1)^{k+l+m} \mathbf{F}_{klm}(\mathbf{r}_\mathrm{s}).
\end{equation}
The $x$-component of $\mathbf{F}$ can be written:
\begin{equation}
    \begin{aligned}
        (F_{klm})_x = &\Delta y_l \log( \Delta z_m + \Delta r_{klm}) + \Delta z_m \log(\Delta y_l
        + \Delta r_{klm})\\
        &- \Delta x_k \arctan \left( \frac{\Delta y_l \Delta z_m}{\Delta x_k \Delta r_{klm}} \right),
    \end{aligned}
\end{equation}
where $\Delta x_k = x_k - x_\mathrm{s}$, $\Delta y_l = y_l - y_\mathrm{s}$, $\Delta z_m = z_m - z_\mathrm{s}$, and $\Delta r_{klm} = (\Delta x_k^2 + \Delta y_l^2 + \Delta z_m^2)^{1/2}$. The $y$ ($z$) component of $\mathbf{F}$ can then be obtained by permuting $\Delta x_k \to \Delta y_l$, $\Delta y_l \to \Delta z_m$, $\Delta z_m \to \Delta x_k$ once (twice).

\begin{figure}[b]
    \centering
    \includegraphics[width=0.5\linewidth]{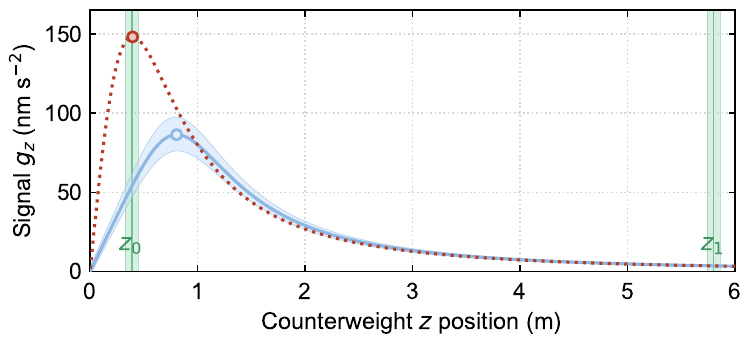}
    \caption{\textbf{Model of counterweight gravitational signal.} Analytical model of the vertical gravitational acceleration $g_z$ due to the counterweight (blue solid line) as a function of the $z$-position of its centre of mass relative to the gravimeter. Shading around the curve denotes one standard deviation error arising from the uncertainty in the counterweight size and density, as well as the distance to the sensor. Vertical lines indicate $z_\mathrm{0}$ and $z_\mathrm{1}$, the lower and upper positions of the counterweight, respectively. The shaded regions show one standard deviation error in these positions. A point mass approximation of the signal is also shown (red dotted line). The maximal signal, corresponding to the position $z_+^*$, is indicated on each line with a circle.}
    \label{fig:model}
\end{figure}

Figure~\ref{fig:model} shows the vertical gravitational signal $g_z$ obtained from the model (solid blue line), with uncertainty indicated by the shading. At large distances, the vertical gravitational force is weak, but increases as the counterweight moves downward. However, past the peak at $\approx 0.8$\,m (identified with a circle), the signal starts to drop towards zero as the gravitational field becomes predominantly along the $x$-axis. This peak corresponds to the optimal position $z_+^*$ (see Supplementary Materials). In Fig.~\ref{fig:model}, we also show the signal predicted for a point mass with the same total mass as the counterweight, located at the position of its COM (dotted line). As expected, the two curves agree in the far-field, where the distributed density begins to look point-like. However, when the counterweight is closer to the sensor, the point mass curve produces a much stronger signal, and the optimal position shifts to $z_+^* \approx 0.4$\,m. This highlights the benefits that could be realised if the mass was made denser (e.g.~compacted into a sphere).

The experimental values for the lower and upper positions of the counterweight ($z_\mathrm{0}$ and $z_\mathrm{1}$, respectively) are indicated in Fig.~\ref{fig:model} with vertical lines (shading denotes the uncertainty in these values). (Details regarding how these positions have been obtained are presented in the following section.) According to the distributed counterweight model, the expected signals at each location are 54(15)\,\nmss\:(lower) and 3.6(3)\,\nmss\:(upper),
giving a difference of $\Delta g_z = 50(15)$\,\nmss. This value agrees well with the experimentally observed signal of $\approx 50$\,\nmss.

\subsection*{Estimation of counterweight geometry and mass} \label{sub:counterweight_geometry}

\begin{figure}[t]
    \centering
    \includegraphics[width=0.5\linewidth]{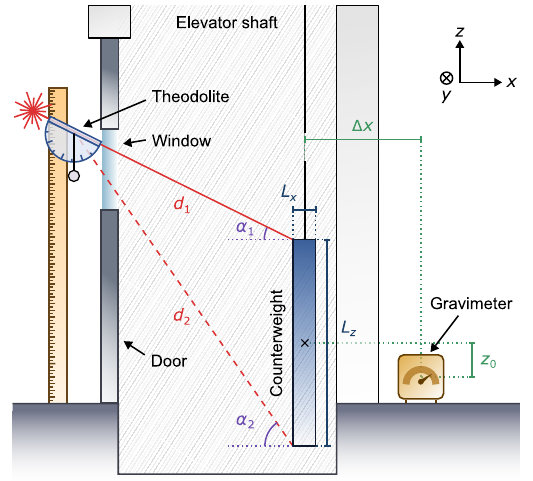}
    \caption{\textbf{Estimation of counterweight geometry.} Schematic diagram of the experimental setup and measurement scheme for the elevator counterweight dimensions while in the lower position (not to scale). The left side shows the combined laser distance meter and theodolite, used for measuring the angle $\alpha_1$ ($\alpha_2$) to the top (bottom) of the counterweight, as well as the distance $d_1$ ($d_2$ could not be measured; see text). 
    The counterweight height $L_z$ and vertical position $z_0$ obtained using this method are indicated, as well as the horizontal depth $L_x$ and distance to the gravimeter $\Delta x$.}
    \label{fig:Sextant}
\end{figure}

To determine the counterweight COM $z$-position relative to the gravimeter in its lower position in the experiment, we used a laser distance meter (LDM) and purpose-built pendulum theodolite, mounted to a tripod (see left side of Fig.~\ref{fig:Sextant}). The pendulum theodolite consisted of a
fishing line with a mass on the end, attached to the centre of a protractor. A laser pointer was also mounted along the edge of the protractor, directed along the 0--$180^\circ$ line. The theodolite was first aligned such that the laser pointed to the top of the counterweight (solid red line in Fig.~\ref{fig:Sextant}), before being rotated until the laser reached the bottom of the counterweight (dashed red line). As it was rotated, the pendulum remained vertical due to gravity, allowing a measurement of the angles $\alpha_1$ and $\alpha_2$ relative to this reference (see Fig.~\ref{fig:Sextant}). We then attached the LDM to the centre of the theodolite, enabling us to measure the distance $d_1$ from this point to the top of the counterweight (the LDM could not provide a reliable measurement of the distance $d_2$ to the bottom of the counterweight, due to poor reflectance). Using these measurements, as well as the measured height of the theodolite, trigonometry was used to determine the counterweight's vertical length, $L_z = 1.47(4)$\,m. 
The counterweight's centre-point (assumed to be its COM) was also estimated from this method to be at a vertical position $z_{0} = 0.39(6)$\,m above the gravimeter (see Fig.~\ref{fig:Sextant}). Building plans informed us that the elevator moves a distance 5.41\,m between the upper and lower levels, and hence the upper position of the counterweight was estimated to be $z_1 = 5.80(6)$\,m.

The depth $L_x$ and width $L_y$ could not be estimated using this method, due to a lack of absolute horizontal reference. Fortuitously, shortly after completing the experiment, a technician performing an elevator service was kind enough to provide us with measurements of $L_x = 0.205(5)$\,m, $L_y = 0.82(1)$\,m. 

To obtain the horizontal distance $\Delta x$ between the counterweight and the gravimeter, we needed an estimate of the wall thickness (the technician also provided a measurement of the distance between the counterweight and the wall). To this end, we conducted a survey with a tape measure and the LDM from our reference point at the front of the elevator shaft to the gravimeter, via a hallway next to the elevator shaft. This gave a wall thickness of $0.27(2)$\,m, consistent with a nearby wall in the same building that we could measure directly. We therefore estimated the total separation to be $\Delta x = 0.56(3)$\,m.

Finally, from the counterweight's measured dimensions, we estimated its mass to be $1800(200)$\,kg, assuming a uniform density of $7300(200)$\,\kgmmm. This density is typical of grey cast iron---the material of choice for mid 20th-century lift counterweights.


\begin{table}[b]
    \centering
    \setlength{\tabcolsep}{4pt}
    \caption{Potential improvements to transmission range.}
    \begin{tabular}{l c c c}
        \hline
        \hline
          & Current & Potential & Range increase factor\\
         \hline
         Geometric factor & 1 & $\approx 4.9$ & $\approx 2.2$  \\ 
         Mass (kg) & 1800 & 75\,000 & $\approx 6.5$ \\
         Sensitivity (nm\,s$^{-2}$) & $\approx 2$ & $\approx 0.01$ & $\approx 14$\\
         SNR & $\approx 25$ & $\approx 2$ & $\approx 3.5$\\
         \hline
         \hline
    \end{tabular}
    \label{tab:range_optimisation}
\end{table}


\clearpage 

\bibliographystyle{sciencemag}
\bibliography{refs.bib}

\paragraph*{Acknowledgments:}
We thank Geoscience Australia for the loan of the gravimeter and for their technical advice, as well as Guillaume Gauthier, Charlotte Thomson, Lewis Williamson and Simon Haine for comments on the manuscript. AGW wishes to thank Paulo Nussenzveig for conversations musing on applications of next generation quantum technologies. Finally, we thank the anonymous referee for drawing our attention to the implicit assumption of a fixed reference frame, and for highlighting that active blocking of the gravitational signal is, in principle, possible.
\paragraph*{Funding:}
AJG, CWW, AGW and MJD were funded by the Australian Research Council Centre of Excellence for Engineered Quantum Systems (EQUS, CE170100009). AJG, MJD and AGW were also funded by the Australian government Department of Industry, Science, and Resources via the Australia-India Strategic Research Fund (AIRXIV000025). AGW is the recipient of an Australian Research Council Laureate Fellowship (FL210100045) funded by the Australian Government.
\paragraph*{Author contributions:}
Conceptualization: CWW, MDH, AJG, AGW, MJD.
Formal analysis: CWW, MDH, AJG, AGW, MJD. 
Investigation: CWW, MDH, AJG, AGW, MJD. 
Methodology: CWW, MDH, AJG, AGW, MJD. 
Supervision: MJD. 
Writing---original draft: CWW, MDH, AJG, AGW, MJD.
\paragraph*{Competing interests:}
The authors declare they have no competing interest.
\paragraph*{Data, code, and materials availability:}
All data and code needed to evaluate and reproduce the results in the paper are available at the University of Queensland eSpace data repository, https://doi.org/10.48610/cd26dd2. This study did not generate new materials.


\subsection*{Supplementary materials}
Supplementary text\\
Figs. S1 to S6\\   

\newpage

\renewcommand{\thefigure}{S\arabic{figure}}
\renewcommand{\thetable}{S\arabic{table}}
\renewcommand{\theequation}{S\arabic{equation}}
\renewcommand{\thepage}{S\arabic{page}}
\setcounter{figure}{0}
\setcounter{table}{0}
\setcounter{equation}{0}
\setcounter{page}{1} 


\begin{center}
\section*{Supplementary Materials for\\ \scititle}

Andrew~J.~Groszek$^{\ast\dagger}$,
Charles~W.~Woffinden$^\dagger$,
Michael~D.~Harvey, 
Andrew.~G.~White, 
Matthew~J.~Davis$^\ast$

\small$^\ast$Corresponding authors. Email: a.groszek@uq.edu.au; mdavis@uq.edu.au\\
\small$^\dagger$These authors contributed equally to this work.
\end{center}

\subsubsection*{This PDF file includes:}
Supplementary text\\
Figs. S1 to S6\\

\newpage



\section{Properties of gravitational communication}

\subsection{Gravitational signal optimisation}\label{sub:optimisation}
In the main text, we discuss the gravitational signal strength $\Delta \mathbf{g} = \mathbf{g}(\mathbf{r}_1) - \mathbf{g}(\mathbf{r}_0)$ that Alice generates as she moves a point mass $M$ between two positions $\mathbf{r}_0$ and $\mathbf{r}_1$ [where $\mathbf{g}(\mathbf{r}) = (GM/r^2)\hat{\mathbf{r}}$]. Here we consider how Alice can optimise her transmission by choosing positions that maximise this signal.

Her choices will depend on several factors. Firstly, many gravimeters only measure along a single axis $\hat{\mathbf{n}}$ (usually the vertical, $\hat{\mathbf{z}}$), in which case Alice should aim to maximise $\Delta g_n \equiv \Delta \mathbf{g} \cdot \hat{\mathbf{n}}$. If, on the other hand, Bob is in possession of a full three-axis sensor, Alice should instead maximise $|\Delta \mathbf{g}|$. 
Secondly, since we are assuming that Alice is confined to a shielded room (see main text), her options for $\mathbf{r}_0$ and $\mathbf{r}_1$ will be naturally constrained. 

As a benchmark, we first consider an idealised case where the mass is free to move anywhere around the gravimeter on the surface of a sphere with radius $R$.
Here, the strongest three-axis signal will be achieved by moving the mass between two diametrically opposite points on the sphere, i.e. $\mathbf{r}_0 = -\mathbf{r}_1$, giving a maximum signal of $|\Delta \mathbf{g}| = 2GM/R^2$. This same signal can also be achieved with a single-axis sensor, provided the positions are also aligned with the measurement axis, i.e.~$\mathbf{r}_0 = -\mathbf{r}_1 = \pm R \hat{\mathbf{n}}$ (if $\hat{\mathbf{n}}$ is rotated from this orientation, the signal will be reduced by a factor of $\hat{\mathbf{r}}_0 \cdot \hat{\mathbf{n}}$).

\begin{figure}[b]
    \centering
    \includegraphics[width=0.8\linewidth]{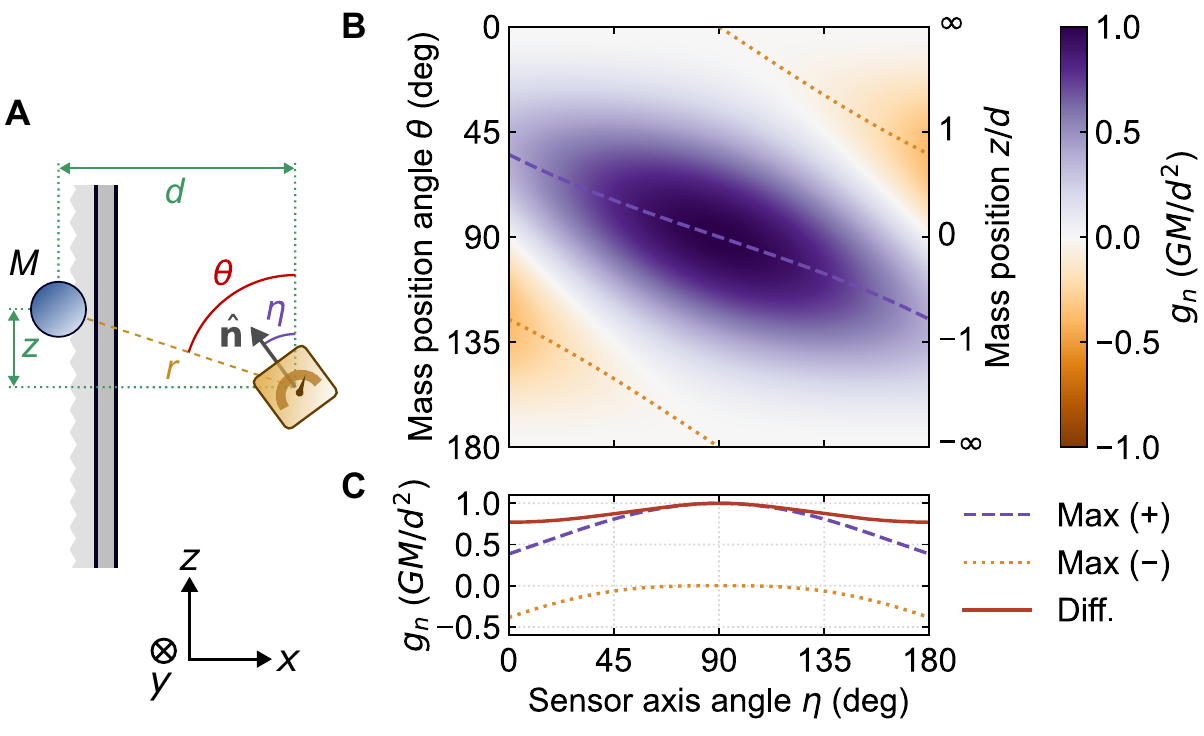}
    \caption{\textbf{Signal optimisation with a vertically constrained mass and a single-axis gravimeter.} (\textbf{A}) Setup with the mass allowed to move vertically on one side of a wall, and the sensor on the other side. The mass position can be parametrised by either its vertical position $z$ or polar angle $\theta$. The sensor's measurement axis $\hat{\mathbf{n}}$ is expressed in terms of a polar angle $\eta$ (measured relative to $\hat{\mathbf{z}}$) and an azimuthal angle $\zeta$ in the $xy$-plane (measured relative to $\hat{\mathbf{x}}$; not shown). (\textbf{B}) The detected gravitational acceleration $g_n(\theta, \eta, \zeta=0)$ [Eq.~\eqref{eq:g_2d}] as a function of sensor axis angle $\eta$ and mass angle $\theta$ (or mass position $z$; right vertical axis). The purple dashed (orange dotted) line shows the location of the maximally positive (negative) value of $g_n$ at each value of $\eta$. (\textbf{C}) The extremal $g_n$ values identified in B, plotted as a function of sensor polar angle $\eta$. The solid red line shows the difference between these extremal values, corresponding to the optimal signal $\Delta g_n$ for a given $\eta$.}
    \label{fig:app_verticalsignal}
\end{figure}

In the scenario envisioned in Fig.~1 of the main text, where a vertical wall separates the mass and the gravimeter, this ideal optimisation strategy is not possible. 
Instead, we consider Alice's mass to be constrained to a vertical line parallel to the wall, a fixed perpendicular distance $d$ from the gravimeter, and with variable vertical position $z$ (see fig.~\ref{fig:app_verticalsignal}A). 
In this configuration, the optimal solution for a three-axis sensor is to move the mass from $z_0=0$ (the closest point to the gravimeter) to $z_1 \to \pm\infty$, giving $|\Delta \mathbf{g}| = GM/d^2$.  
This is a factor of two worse than the ideal case above (taking $R = d$), owing to the restriction that the mass remains on one side of the sensor.

For a single-axis sensor, the situation is more complicated. In the remainder of this section, we wish to determine the positions $z_0$ and $z_1$ that maximise $\Delta g_n$ for a given sensor direction $\hat{\mathbf{n}}$. Trivially, the strongest signal will be obtained when $\hat{\mathbf{n}}$ points directly towards the wall (along the direction $\hat{\mathbf{x}}$ depicted in fig.~\ref{fig:app_verticalsignal}A). In this case, the optimal mass positions and signal are the same as for the three-axis sensor. However, since most sensors measure vertical gravity (i.e.~$\hat{\mathbf{n}} = \hat{\mathbf{z}}$), the natural question is how much sensitivity is lost as $\hat{\mathbf{n}}$ is rotated from the horizontal ($\hat{\mathbf{x}}$) to the vertical ($\hat{\mathbf{z}}$) axis.

We first express the vertical position of the mass in terms of its polar angle $\theta$ from the sensor (related to its height via $z = d/\tan\theta$). It is convenient to parametrise $\hat{\mathbf{n}}$ in terms of a polar angle $0^\circ \leq \eta < 180^\circ$ (as shown in fig.~\ref{fig:app_verticalsignal}A) and an azimuthal angle $0^\circ \leq \zeta < 360^\circ$ (the rotation around $\hat{\mathbf{z}}$, which we include for generality). The detected gravitational pull towards the mass, $g_n = \mathbf{g} \cdot \hat{\mathbf{n}}$, can then be written in terms of these three angles:
\begin{equation} \label{eq:g_2d}
    g_n(\theta, \eta, \zeta) = \frac{GM}{d^2} \sin^2\theta \left( \cos\theta \cos\eta + \sin\theta \sin\eta \cos\zeta \right).
\end{equation}
This function is plotted in terms of $\theta$ and $\eta$ in fig.~\ref{fig:app_verticalsignal}B for $\zeta=0$. For a given sensor angle $\eta$, there are two extremal values of the gravitational signal---a maximally positive and a maximally negative value. These occur at mass angles 
\begin{equation} \label{eq:theta_star}
\theta^*_\pm(\eta) = \eta/2 \pm \arccos(-\cos(\eta)/3)/2
\end{equation}
(modulo $180^\circ$). Here, the $^*$ notation identifies these as extremal points, and $\pm$ denotes the maximally positive ($+$) and maximally negative ($-$) position. These extrema are identified in fig.~\ref{fig:app_verticalsignal}B with a purple dashed and orange dotted line, respectively. Note that, because we have assumed the sensor always points towards the wall, the maximally positive signal is larger in magnitude than the maximally negative signal for all $\eta$. If we instead tilted the sensor away from the wall by a polar angle $\eta$, the situation would be reversed.

Figure~\ref{fig:app_verticalsignal}C shows these extremal values as a function of $\eta$ (using the same line types). Also shown in this frame is the difference between the two curves, corresponding to the largest attainable value of $\Delta g_n$ for a given sensor angle $\eta$ (solid red line). In practice, this is obtained by choosing mass positions $\theta_0 = \theta^*_\pm$ and $\theta_1 = \theta^*_\mp$. Evidently, the global maximum of the signal is obtained when $\eta=90^\circ$ (i.e.~the sensor pointing along $\hat{\mathbf{x}}$, as identified earlier). In this case, the optimal mass positions are $\theta^*_+ = 90^\circ$ and $\theta^*_- = 0^\circ$ or $180^\circ$, as identified above (see fig.~\ref{fig:app_verticalsignal}B). As the sensor is rotated towards $\hat{\mathbf{z}}$, the optimal signal reduces, until it reaches a minimum at $\eta = 0^\circ$ (or equivalently, $\eta = 180^\circ$). Remarkably, though, this signal is only $\approx 23\%$ smaller than the global maximum (i.e.~$\approx 0.77 GM/d^2$), despite the sensor now pointing parallel to the wall. In this case, the optimal choice of mass angles is $\theta^*_+ \approx 55^\circ$, $\theta^*_- \approx 125^\circ$ [from Eq.~\eqref{eq:theta_star}]---or equivalently, $z^*_\pm \approx \pm 0.7 d$ (as stated in the main text). We therefore conclude that the measurement axis of the sensor makes only a modest difference to the overall sensitivity achievable in this configuration.

However, we note that the movement distance of $\approx$~$1.4 d$ required to achieve this optimal signal is larger than the transmission length $d$, which may be unrealistic in practice. Taking instead the limit of a small mass movement $|z_0| \ll d$ symmetric around $z=0$ (i.e.~$z_0 = -z_1$), Eq.~\eqref{eq:g_2d} gives $\Delta g_z \approx (2z_0/d)(GM/d^2)$ (still taking $\eta = 0^\circ$). The signal in this limit will therefore be a factor of $2z_0/d$ smaller than the global maximum.

For completeness, we also consider the effect of nonzero azimuthal angle $\zeta$ (rotating $\hat{\mathbf{n}}$ into or out of the page in fig.~\ref{fig:app_verticalsignal}A). This can only ever reduce the signal, since the mass is confined to the $xz$-plane, and should therefore be avoided. As $\zeta$ increases from $0^\circ$, the central peak in $\Delta g_n$ (solid line in fig.~\ref{fig:app_verticalsignal}C) reduces in value, eventually becoming a dip, while the endpoints at $\eta = 0^\circ$ and $\eta = 180^\circ$ remain fixed (because the azimuthal angle is irrelevant when the sensor is pointing along $\hat{\mathbf{z}}$). For $\zeta = \eta = 90^\circ$, the signal vanishes completely, because the sensor is pointing along $\hat{\mathbf{y}}$, and hence it will always be perpendicular to the mass.

\subsection{Derivation of antenna signal}\label{app:antenna_derivation}

\begin{figure}[t]
    \centering
    \includegraphics[width=0.8\linewidth]{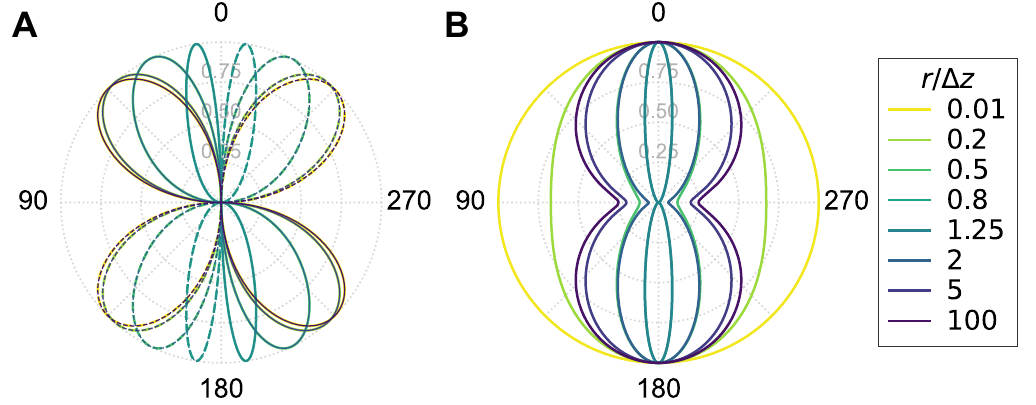}
    \caption{\textbf{Gravitational dipole antenna signal.} Polar plots of the normalised signal from the gravitational antenna configuration shown in Fig.~2A of the main text, as a function of polar angle $\alpha$ from the axis of the antenna. Here we show (\textbf{A}) the horizontal signal $\Delta g_x$, and (\textbf{B}) the total signal $|\Delta \mathbf{g}|$ (the vertical signal $\Delta g_z$ is shown in Fig.~2B of the main text). The legend denotes the value of $r/\Delta z$ for each curve, and each curve is normalised to the largest value for that radius. In A, solid (dashed) lines denote a positive (negative) signal. Note that all patterns are symmetric with respect to the observer's azimuthal angle.}
    \label{fig:antenna_x_total}
\end{figure}

In the main text, we consider the transmission pattern produced by a gravitational antenna, formed by moving a point mass $M$ between two vertical positions $+\Delta z$ and $-\Delta z$ (see Fig.~2A). An observer at position $\mathbf{r}$ will detect a change in gravitational acceleration as the mass is moved between the upper (1) and lower (0) positions, given by:
\begin{equation} \label{Eq:delta_g}
    \Delta \mathbf{g} (\mathbf{r}) = \mathbf{g}_\mathrm{1}(\mathbf{r}) - \mathbf{g}_\mathrm{0}(\mathbf{r}) = GM \left[ \frac{\mathbf{r}_\mathrm{1} - \mathbf{r}}{|\mathbf{r}_\mathrm{1} - \mathbf{r}|^3} - \frac{\mathbf{r}_\mathrm{0} - \mathbf{r}}{|\mathbf{r}_\mathrm{0} - \mathbf{r}|^3}\right],
\end{equation} 
where $\mathbf{r}_\mathrm{1} = \Delta z \hat{\mathbf{z}} = -\mathbf{r}_\mathrm{0}$. This signal is confined to the 2D plane containing $\textbf{r}_0$, $\textbf{r}_1$ and $\textbf{r}$, which we refer to as the $xz$-plane without loss of generality. The position of the gravimeter in this plane can then be expressed in polar coordinates as $\mathbf{r}=(r, \alpha)$, in which case the horizontal and vertical components of the above expression are, respectively:
\begin{equation} \label{eq:antenna_components}
\begin{aligned}
    \Delta g_x(\mathbf{r}) &= GM \left( -\frac{r\sin\alpha}{\delta_+^3} + \frac{r\sin\alpha}{\delta_-^3} \right), \\
    \Delta g_z(\mathbf{r}) &= GM \left( \frac{\Delta z - r\cos\alpha}{\delta_+^3}
    + \frac{\Delta z + r\cos\alpha}{\delta_-^3} \right),
\end{aligned}
\end{equation}
where $\delta_\pm = (r^2 + \Delta z^2 \pm 2 \Delta z r \cos\alpha)^{1/2}$. Figure~\ref{fig:antenna_x_total} presents normalised polar plots of both the horizontal signal $\Delta g_x$ (panel A) and the magnitude of the total signal $|\Delta \mathbf{g}| = (\Delta g_x^2 + \Delta g_z^2)^{1/2}$ (panel B) as a function of polar angle $\alpha$, at multiple values of $r/\Delta z$ (the vertical signal $\Delta g_z$ is shown in Fig.~2B of the main text). In the near- and far-field limits ($r \ll \Delta z$ and $r \gg \Delta z$, respectively), $\Delta g_x$ consists of four lobes, centred at angles $\alpha = \lbrace 45^\circ, 135^\circ,225^\circ, 315^\circ \rbrace$. This horizontal signal is unchanged upon exchanging $r \leftrightarrow \Delta z$, up to a rescaling by $r/\Delta z$ [see Eq.~\eqref{eq:antenna_components}]; as such, the normalised plots of $\Delta g_x$ are symmetric about $r/\Delta z = 1$. 
An observer in the far-field and an observer near the origin therefore see equivalent $\Delta g_x$ patterns. However, the $\Delta g_x$ component is small in magnitude compared to $\Delta g_z$, and therefore has only a minor impact on the shape of the overall signal $|\Delta \mathbf{g}|$. This is evident from fig.~\ref{fig:antenna_x_total}B, which looks qualitatively similar the plot of $\Delta g_z$ in Fig.~2B.

Taking the far-field limit ($r \gg \Delta z$), Eqs.~\eqref{eq:antenna_components} reduce to:
\begin{equation} \label{eq:antenna_farfield}
\begin{aligned}
    \Delta g_x &\approx 6\Delta z  \frac{GM}{r^3} \sin(\alpha) \cos(\alpha),\\
    \Delta g_z &\approx 6\Delta z \frac{GM}{r^3} \left(\cos^2(\alpha) + \frac{1}{3} \right).
\end{aligned}
\end{equation}
From these expressions, the far-field half-power beam widths (HPBWs) of each component can be determined, which characterise the angular spread of the signal. These are: $45^\circ$ (horizontal, $\Delta g_x$), $2\arccos(1/3^{1/2}) \approx 109^\circ$ (vertical, $\Delta g_z$), and $\arccos(-3/5) \approx 127^\circ$ (total, $|\Delta \mathbf{g}|$). Therefore, measuring $|\Delta \mathbf{g}|$, rather than $\Delta g_z$, serves to widen the antenna lobes slightly in the far-field limit, as can be seen by comparing fig.~\ref{fig:antenna_x_total}B and Fig.~2B.
The $\Delta g_z$ and $|\Delta \mathbf{g}|$ HPBWs are somewhat greater than the $\approx 78^\circ$ of the electromagnetic half-wave dipole antenna, whose radiation pattern is given by $\sim$~$\cos^2[(\pi/2)\cos \alpha ]/\sin^2 \alpha$~\cite{ramo1994} (see Fig.~2B). The gravitational antenna would therefore typically be considered less efficient, as it transmits signal over a wider range, including partially in the perpendicular direction.

The equivalent electrostatic situation, obtained by replacing the mass $M$ with a charge $Q$, and the acceleration field $\mathbf{g}$ with the electric field $\mathbf{E}$, would produce identical antenna properties to those described above, up to an additional replacement of $G \to -1/4\pi\varepsilon_0$ in Eqs.~\eqref{Eq:delta_g}--\eqref{eq:antenna_farfield} (with $\varepsilon_0$ the vacuum permittivity).

\subsection{Robustness to signal jamming, blocking and spoofing}

Here we consider attacks on Alice's gravitational signal by a malicious actor, Mallory. These can come in the form of jamming (overwhelming the signal at Bob's gravimeter with other gravitational effects), blocking (cancelling Alice's signal to prevent it from being detectable by Bob's gravimeter), and spoofing (producing the same signal at Bob's gravimeter as Alice, allowing nefarious data transmission). We begin in sec.~\ref{sub:single_mass} by considering the scenario where Mallory, like Alice, controls a single mass. Here, gravitational jamming is relatively straightforward, whereas blocking and spoofing are not. To realise true blocking and spoofing, Mallory must have access to a surface of infinitely configurable mass density akin to an electromagnetic Faraday cage---a hypothetical we explore in sec.~\ref{sub:configurable_mass}.

An important consideration in this analysis is the specific form of gravitational information accessible to Bob. With a three-axis gravimeter, he will in general measure the gravitational vectors $\mathbf{g}_0 + \mathbf{g}_\mathrm{bg}$ and $\mathbf{g}_1 + \mathbf{g}_\mathrm{bg}$ in Alice's 0 and 1 positions, respectively. Here, $\mathbf{g}_\mathrm{bg}$ is the background signal arising from genuine gravitational effects other than Alice (such as the pull of the Earth, Moon and Sun, as well as Mallory's mass), plus any systematic sensor offsets such as drift and---in the case that Bob is using a relative gravimeter---an arbitrary constant offset. If Bob is able to measure $\mathbf{g}_\mathrm{bg}$ directly (e.g.~by performing a background measurement without Alice present), he can subtract it to isolate the true $\mathbf{g}_0$ and $\mathbf{g}_1$ vectors independently, providing him with additional information to use against attacks. For a single-axis sensor measuring along direction $\hat{\mathbf{n}}$, Bob will only detect $(\mathbf{g}_0 + \mathbf{g}_\mathrm{bg}) \cdot \hat{\mathbf{n}}$, $(\mathbf{g}_1 + \mathbf{g}_\mathrm{bg}) \cdot \hat{\mathbf{n}}$, and potentially a background measurement $\mathbf{g}_\mathrm{bg} \cdot \hat{\mathbf{n}}$.

\subsubsection{Malicious attacks with a single mass \label{sub:single_mass}}

If Mallory moves a single mass $\tilde{M}$ nearby to Bob, they could create a time varying contribution to the background field to overwhelm Alice's true signal and jam the communication channel. With a single-axis sensor, Bob is particularly vulnerable to this type of attack, as he would not be able to use directional information to identify which part of the signal is coming from Alice's true direction. A three-axis sensor, on the other hand, would detect the jamming signal in all three vector components, and this could provide valuable additional information. For example, if Alice's mass is known to be moving only in a single axis, the malicious signal in the perpendicular components would be correlated with the noise in the direction of the true mass, allowing Alice to isolate this signal and remove it. In more general situations, isolating the true signal becomes a nontrivial inverse problem, although it may be solvable if enough information about Alice's true signal is known (e.g.~the expected amplitude or frequency).

\begin{figure}[b]
    \centering
    \includegraphics[width=0.9\linewidth]{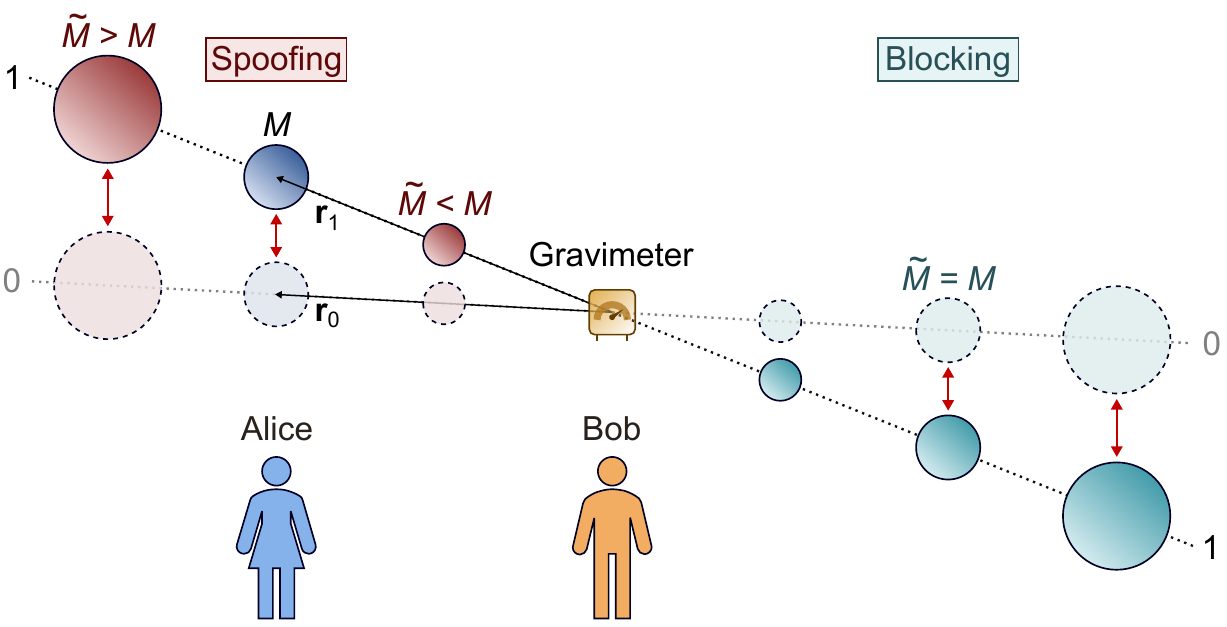}
    \caption{\textbf{Gravitational signal spoofing and blocking with a single mass.} Schematic demonstrating Mallory's attempts to spoof or block Alice's signal using a second point-like mass $\tilde{M}$. Alice moves her mass $M$ between two positions $\mathbf{r}_i$, with $i \in \lbrace 0, 1 \rbrace$, to produce signals $\mathbf{g}_i$ at Bob's gravimeter. To spoof ($+$) or block ($-$) the signal, Mallory must position their mass at $\tilde{\mathbf{r}}_i = \pm (\tilde{M}/M)^{1/2} \mathbf{r}_i$, which correspond to the left and right side of the figure, respectively. Mallory can be either closer to Bob than Alice (if $\tilde{M} < M$) or further away (if $\tilde{M} > M$).}
    \label{fig:mallory}
\end{figure}

To block ($-$) or spoof ($+$) Alice's signal, Mallory must position their mass $\tilde{M}$ so as to produce signals $\tilde{\mathbf{g}}_i = \pm \mathbf{g}_i$ at Bob's gravimeter, where the index $i \in \lbrace 0, 1 \rbrace$ denotes Alice's 0 and 1 positions. The only positions satisfying this are $\tilde{\mathbf{r}}_i = \pm (\tilde{M}/M)^{1/2} \, \mathbf{r}_i$ (with Bob's gravimeter taken to be the coordinate origin). For spoofing (blocking), Mallory's mass must be therefore be placed on the same (opposite) side of the sensor to Alice, as depicted in fig.~\ref{fig:mallory}. If Bob cannot measure the background $\mathbf{g}_\mathrm{bg}$, then Mallory is less constrained. Here, only the change in Alice's signal $\Delta \mathbf{g} = \mathbf{g}_1 - \mathbf{g}_0$ must be blocked or spoofed, which can be achieved by ensuring that 
\begin{equation} \label{Eq:delta_g_spoofing}
    \Delta \mathbf{g} (\mathbf{r}) = M \left[ \frac{\mathbf{r}_\mathrm{1}}{|\mathbf{r}_\mathrm{1}|^3} - \frac{\mathbf{r}_\mathrm{0}}{|\mathbf{r}_\mathrm{0}|^3}\right] = \pm \tilde{M} \left[ \frac{\tilde{\mathbf{r}}_\mathrm{1}}{|\tilde{\mathbf{r}}_\mathrm{1}|^3} - \frac{\tilde{\mathbf{r}}_\mathrm{0}}{|\tilde{\mathbf{r}}_\mathrm{0}|^3}\right].
\end{equation}
For example, Mallory could choose an arbitrary initial position $\tilde{\mathbf{r}}_0$, and then find the position $\tilde{\mathbf{r}}_1$ that satisfies the above for their mass $\tilde{M}$.

If Bob's sensor only measures along a single direction $\hat{\mathbf{n}}$, Mallory only needs to satisfy $\mathbf{g}_i \cdot \hat{\mathbf{n}} = \pm \tilde{\mathbf{g}}_i \cdot \hat{\mathbf{n}}$, or equivalently $(M/r_i^2) \cos(\varphi_i) = \pm (\tilde{M}/\tilde{r}_i^2) \cos(\tilde{\varphi}_i)$, where $\varphi_i$ and $\tilde{\varphi}_i$ are the angles between $\hat{\mathbf{n}}$ and $\mathbf{r}_i$, and $\hat{\mathbf{n}}$ and $\tilde{\mathbf{r}}_i$, respectively. This gives Mallory additional freedom to trade off between distance $\tilde{r}_i$ and angle $\tilde{\varphi}_i$ while maintaining the same signal. These constraints can be relaxed even further if Bob does not know $\mathbf{g}_\mathrm{bg} \cdot \hat{\mathbf{n}}$, in which case Mallory only needs to satisfy $\Delta \mathbf{g} \cdot \hat{\mathbf{n}} = \pm \Delta \tilde{\mathbf{g}} \cdot \hat{\mathbf{n}}$.

\subsubsection{Malicious attacks with a continuous mass distribution \label{sub:configurable_mass}}

In the above analysis, the signal is only blocked or spoofed at a single location, so Bob can defend against Mallory's attacks with a single mass by simply repositioning himself, forcing Mallory to find new positions $\tilde{\mathbf{r}}_0$ and $\tilde{\mathbf{r}}_1$. To achieve true blocking and spoofing throughout an extended region of space, Mallory would instead need to construct an effective Faraday cage for gravity. While completely impractical, we explore this scenario here for theoretical completeness. As the situation is largely analogous to electrostatics, we begin by summarising the physics of electromagnetic Faraday shielding.

\begin{figure}[t]
    \centering
    \includegraphics[width=0.8\linewidth]{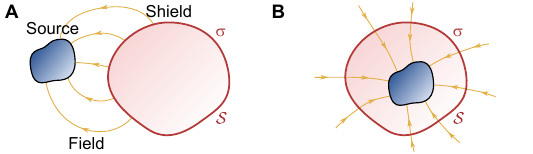}
    \caption{\textbf{Faraday shielding of an electric or gravitational field.} A source charge distribution (either electric charge or gravitational mass) produces a field ($\mathbf{E}$ or $\mathbf{g}$, respectively) incident on a closed surface $\mathcal{S}$ carrying areal charge density $\sigma(\mathbf{r})$. If the source field is incident from outside $\mathcal{S}$ (\textbf{A}), complete blocking of the field is possible inside the surface. If the source is instead contained inside $\mathcal{S}$ (\textbf{B}), the field outside will be nonzero in general, although its spatial configuration will not depend on the source distribution---only on the geometry and total charge of the surface. In this case, the field outside can only be cancelled if the source and surface charges add to zero, which is not possible for gravity, since mass can only be positive.}
    \label{fig:blocking}
\end{figure}

\newpage
\noindent\textbf{Electromagnetic shielding}\\
A closed surface $\mathcal{S}$ of perfectly conducting material will act as a Faraday cage, automatically shielding any electric field $\mathbf{E}_\mathrm{source}$ incident on it from outside~\cite{griffiths_introduction_2023}. Physically, the areal electric charge density $\sigma(\mathbf{r})$ on the surface of the conductor rearranges itself into a configuration that produces its own field $\mathbf{E}_\mathrm{conductor}$ that perfectly counteracts the incident field (see fig.~\ref{fig:blocking}A). In terms of electric potential $V(\mathbf{r})$, this charge redistribution enforces the condition $V_\mathrm{source}(\mathbf{r}) + V_\mathrm{conductor}(\mathbf{r}) = V_\mathcal{S}$ at the surface of the conductor, with $V_\mathcal{S}$ a constant. Throughout the volume inside the conducting surface, Poisson's equation $\nabla^2 V = 0$ holds, and by its uniqueness theorem, a constant potential at the boundary ensures the same constant value throughout this volume. Since the electric field $\mathbf{E} = -\nabla V$, this in turn ensures zero electric field inside the conductor.

Mathematically, the above condition of constant potential at the surface enforces a charge density that satisfies:
\begin{equation} \label{eq:fredholm}
    \frac{1}{4\pi\varepsilon_0} \int_{\mathcal{S}} \frac{\sigma(\mathbf{r}')}{|\mathbf{r} - \mathbf{r}'|} \mathrm{d}S' = V_\mathcal{S} - V_\mathrm{source}(\mathbf{r}), \hspace{1em} \mathbf{r} \in \mathcal{S}.
\end{equation}
This integral equation can be solved uniquely for $\sigma(\mathbf{r})$ given a particular surface geometry, source potential, and constant $V_\mathcal{S}$. Importantly, shielding occurs inside $\mathcal{S}$ regardless of the net charge $Q_\mathrm{conductor} = \int_{\mathcal{S}} \sigma(\mathbf{r}) \mathrm{d}S$ on the conductor, although this constraint does alter the solution $\sigma(\mathbf{r})$ to Eq.~\eqref{eq:fredholm} (via a change in $V_\mathcal{S}$). In principle, it is possible to find a solution with $\sigma(\mathbf{r}) \geq 0$ everywhere on $\mathcal{S}$, provided $Q_\mathrm{conductor}$ is allowed to be arbitrarily large (this consideration will be important in the gravitational case, since mass density must be positive).

In the inverted scenario, where the field $\mathbf{E}_\mathrm{source}$ is produced by a charge distribution \textit{inside} the conducting surface (fig.~\ref{fig:blocking}B), the conductor will likewise rearrange its charge to make its surface equipotential. However, in this case, the electric field will not only be nontrivial inside the surface, but also outside, in general. By Gauss' law, any net charge $Q_\mathrm{source} + Q_\mathrm{conductor}$ will be detectable by an observer outside the surface, and only if $Q_\mathrm{source} = -Q_\mathrm{conductor}$ will the field outside exactly vanish. Perfect screening in this case therefore requires the existence of negative charge (unless both charge distributions are trivially zero everywhere, in which case no field exists). Nonetheless, all spatial information about the source charge distribution is successfully screened: while the observer outside the surface can in general detect the \textit{presence} of the source charge $Q_\mathrm{source}$ within $\mathcal{S}$, they cannot tell the difference between two source distributions with the same total charge.

In summary, complete Faraday shielding of an electric field is possible with a closed, perfectly conducting surface:
\begin{itemize}
    \item For a $V_\mathrm{source}(\mathbf{r})$ incident from outside the surface, even with the constraint $\sigma(\mathbf{r}) \geq 0$.
    \item For a $V_\mathrm{source}(\mathbf{r})$ incident from inside the surface, but only if charge is allowed to be negative. Regardless of sign constraints, spatial information about the source distribution is lost outside the surface.
\end{itemize}

\vspace{0.5\baselineskip}
\noindent\textbf{Gravitational shielding}\\
We now consider the possibility of gravitational Faraday shielding. The gravitational source potential produced by Alice's mass distribution $\rho(\mathbf{r})$ (in either the 0 or 1 configuration) is
\begin{equation}
\Phi_\mathrm{source}(\mathbf{r}) = - G \int_{\mathcal{V}} \frac{\rho(\mathbf{r}')}{|\mathbf{r} - \mathbf{r}'|} \mathrm{d}^3 \mathbf{r}',
\end{equation}
where $\mathcal{V}$ is the region containing the nonzero mass, and Alice's total mass is $M = \int_\mathcal{V} \rho(\mathbf{r}) \mathrm{d}^3 \mathbf{r}$. Bob is trying to detect the gravitational field $\mathbf{g} = -\nabla \Phi_\mathrm{source}$ produced by this mass distribution at a nearby location. The question we wish to answer is whether this field can be blocked everywhere in Bob's region of space.

There are two major points of difference between this scenario and the electrostatic case described above: (i) while electric charge can take both signs, mass can only be positive; (ii) there is no equivalent of a `conductor' for mass---a material that, with no external input, rearranges itself to enforce constant gravitational potential along its surface. The second point ensures that gravitational shielding cannot be passive, and must instead be \textit{actively} implemented, presumably by our malicious actor, Mallory. In analogy with electrostatics, Mallory can actively block Alice's signal by creating a closed surface $\tilde{\mathcal{S}}$ of configurable surface mass density $\tilde{\sigma}(\mathbf{r})$ and total mass $\tilde{M} = \int_{\tilde{\mathcal{S}}} \tilde{\sigma}(\mathbf{r}) \mathrm{d}S$.

Once again, there are two possible geometries. In the first, Mallory's surface can surround Bob, letting $\Phi_\mathrm{source}(\mathbf{r})$ be incident from outside (fig.~\ref{fig:blocking}A). Here, Mallory can find a mass density $\tilde{\sigma}(\mathbf{r}) \geq 0$ that enforces a constant $\Phi(\mathbf{r}) = \Phi_{\tilde{\mathcal{S}}}$ everywhere at the surface, thus giving $\mathbf{g}=\mathbf{0}$ inside. In analogy with Eq.~\eqref{eq:fredholm}, this is the solution to
\begin{equation} \label{eq:fredholm_gravity}
    -G \int_{\tilde{\mathcal{S}}} \frac{\tilde{\sigma}(\mathbf{r}')}{|\mathbf{r} - \mathbf{r}'|}  \mathrm{d}S' = \Phi_{\tilde{\mathcal{S}}} - \Phi_\mathrm{source}(\mathbf{r}), \hspace{1em} \mathbf{r} \in \tilde{\mathcal{S}}.
\end{equation}
Alternatively, Mallory can surround Alice, resulting in a field $\Phi_\mathrm{source}(\mathbf{r})$ incident from inside the surface (fig.~\ref{fig:blocking}B). Here there is a difference to electrostatics: since mass can only be positive, the \textit{presence} of Alice's mass can never be hidden by Mallory. (From Gauss' law, Bob can always detect the total mass $M + \tilde{M}$, so Mallory's presence will likewise be detectable.) In this sense, gravity is truly unblockable where electromagnetism is not. However, as in electrostatics, any spatial information about Alice's mass distribution can be shielded outside $\tilde{\mathcal{S}}$ by making the surface an equipotential. This is an important factor for our communication scheme, as Bob can not know whether the mass is in the 0 or the 1 configuration. In fact, if Mallory only needs to block the change in gravity $\Delta \mathbf{g} = \mathbf{g}_1 - \mathbf{g}_0$ in Bob's region of space (e.g.~because Bob cannot measure $\mathbf{g}_\mathrm{bg}$), the equipotential constraint $\Phi(\mathbf{r})=\Phi_{\tilde{\mathcal{S}}}$ can be relaxed in both of the above scenarios. Here, Mallory would only need to reconfigure the surface to have the same total potential $\Phi_0(\mathbf{r}) = \Phi_1(\mathbf{r})$ in Alice's two configurations, as this would give $\Delta \mathbf{g} = \mathbf{0}$ everywhere inside $\tilde{\mathcal{S}}$ for a source distribution outside (and vice versa).

Of course, for any of this to be possible, Mallory must (i) surround either Alice or Bob (but not both) with the surface $\tilde{\mathcal{S}}$, (ii) measure $\Phi_\mathrm{source}$ along this surface, presumably using an array of gravimeters, (iii) use Eq.~\eqref{eq:fredholm_gravity} to determine the required surface density $\tilde{\sigma}(\mathbf{r})$ needed to realise the required potential along $\tilde{\mathcal{S}}$, and (iv) continuously reconfigure this surface density accordingly. This already places substantial demands on Mallory. Moreover, these requirements must be met on the same timescale as Alice's mass motion; otherwise, Bob will have sufficient time to detect the signal. We therefore conclude that, although possible in theory, true gravitational blocking is completely impractical.

The analysis for signal spoofing is nearly identical, except that Mallory should instead configure $\tilde{\sigma}(\mathbf{r})$ to reproduce Alice's potential on $\tilde{\mathcal{S}}$, $\Phi(\mathbf{r}) = \Phi_{\mathrm{source}}(\mathbf{r})$, rather than cancel it. The uniqueness theorem for Laplace's equation then implies that Mallory's mass distribution would generate the same potential as Alice throughout the mass-free region (inside or outside the surface), making the spoofed signal indistinguishable from the genuine one. True signal spoofing across an extended region of space is therefore, like blocking, theoretically possible but essentially impossible in practice.

A more manageable approach to blocking and spoofing would be for Mallory to arrange a set of point masses on $\tilde{\mathcal{S}}$ so as to reproduce the successive multipole moments of $\Phi_\mathrm{source}(\mathbf{r})$ (starting with the dipole, then the quadrupole, and so on). This would be particularly effective in the far-field limit (i.e.~far from both Alice and Mallory's mass distributions) where the dipole term dominates, as Mallory could spoof (block) most of Alice’s signal simply by generating a mass dipole of equal magnitude and (anti)parallel orientation.

Ultimately, the non-negativity of mass ensures that, in all of the above scenarios, Bob will observe a change in background signal $\mathbf{g}_\mathrm{bg}$ when Mallory's mass enters or leaves the configuration, as its presence cannot be concealed gravitationally. Mallory can therefore only ensure secrecy by arriving before Bob and leaving after him---otherwise, Bob will be aware of the intervention. (The exception to this is if Bob has a single-axis gravimeter, in which case Mallory can enter or leave the configuration in the plane perpendicular to $\hat{\mathbf{n}}$.) Given that this communication scheme is restricted to relatively short range applications, Mallory (along with their mass distribution) may nevertheless find it challenging to remain unnoticed during these attempts.

\section{Experimental considerations}

\subsection{Raw gravitational signal \label{sub:raw_gz_data}}

\begin{figure*}[t!]
    \centering
    \includegraphics[width=1\linewidth]{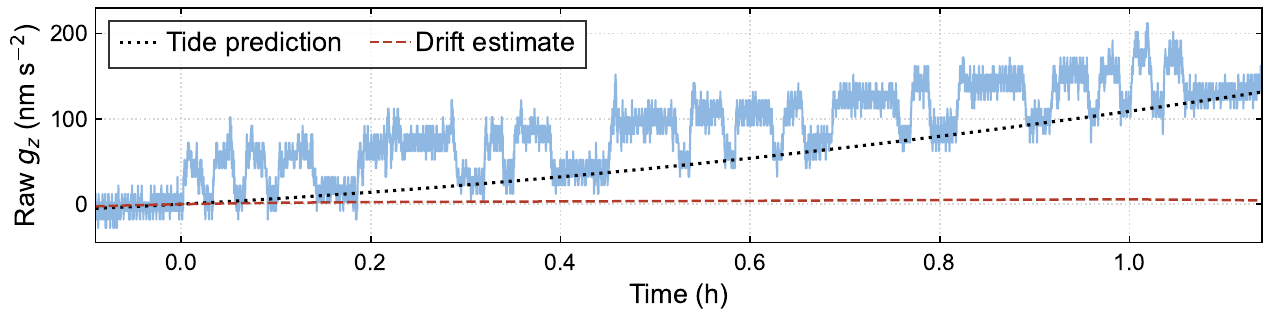}
    \caption{\textbf{Raw experimental gravitational signal.} Raw gravimetry signal $g_z$ over the transmission sequence (solid line), as well as the Longman tide prediction~\cite{Longman_1959} (black dotted line) and an estimate of the drift (red dashed line). Note that the gravimeter outputs a relative measurement, meaning that the overall offset of the data is arbitrary. Note also the sign convention of the sensor, whereby a positive change in $g_z$ corresponds to an increase in downward acceleration.}
    \label{fig:grav_data_raw}
\end{figure*}

The raw vertical gravitational signal $g_z$ detected by the gravimeter over the experimental sequence is plotted in fig.~\ref{fig:grav_data_raw}. The dotted black line shows the Longman Earth tide prediction~\cite{Longman_1959}, which captures the majority of the signal drift over time. We also account for sensor drift and any second order tidal effects with an additional quartic fit (red dashed line) to the data after subtracting the tide prediction. This fit includes the background measurements before and after the bit sequence, as well as the two strings of $\geq 3$ consecutive zeros within the sequence. Subtracting these two contributions from the raw data give the corrected data presented in Fig.~4 of the main text.

\subsection{Insensitivity to magnetic fields} \label{app:magnetic_fields}
Because the gravimeter uses a steel spring, measurements could in principle be influenced by magnetic fields. To mitigate this, the sensor contains two layers of $\mu$-metal shielding~\cite{landr}. However, this shielding is only effective up to a saturation point, and provides a finite attenuation factor, rather than complete suppression. Sufficiently strong fields could therefore still give rise to erroneous gravitational measurements.

To verify that the experimentally observed signal was not due to the iron counterweight's magnetic field, we repeated the experiment with an additional layer of magnetic shielding around the gravimeter. A magnetometer mounted on the side of the gravimeter confirmed that this additional shielding reduced the residual magnetic field by a factor of $\sim 10$. However, when the counterweight was moved between its upper and lower positions with this shielding in place, the obtained gravitational signal was unchanged. We therefore conclude that the measured signal is indeed gravitational in origin.

\subsection{Allan deviation}\label{app:allan_deviation}

\begin{figure}[b]
    \centering
    \includegraphics[width=0.65\linewidth]{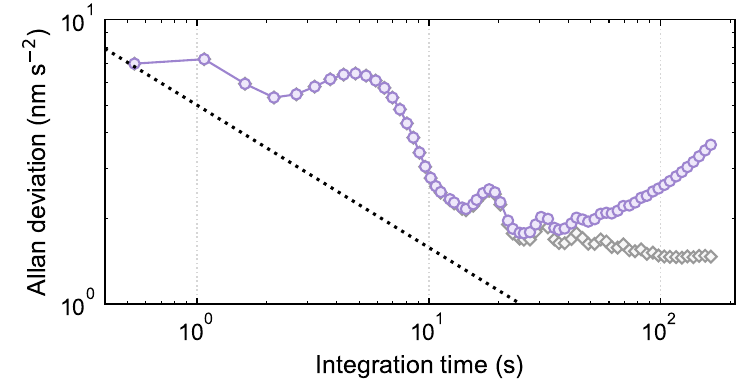}
    \caption{\textbf{Sensor Allan deviation.} Calculated from the $\approx$ 25 minutes of background measurement obtained before the experimental transmission sequence. Purple circles (grey diamonds) show the Allan deviation calculated using the raw data (tide- and drift-corrected data). The dotted line shows a $t^{-1/2}$ power-law for comparison.}
    \label{fig:allan_dev}
\end{figure}

Figure~\ref{fig:allan_dev} shows the Allan deviation of the gravimeter, calculated from the $\approx 25$ minutes of background reading obtained before beginning the experimental sequence. Integrating using the raw data (circles) gives a minimum uncertainty of $\approx$~$2$\,\nmss\:at $\approx$ 30\,s, before this value begins to increase again due to the influence of tides and drift. If we instead calculate the Allan deviation of the tide- and drift-corrected signal over the same window (diamonds), the uncertainty continues to decrease, and appears to reach a minimum after $\approx 100$\,s of integration, with a value of $\approx$~$1.4$\,\nmss. Both curves appear to roughly follow $\sim t^{-1/2}$ scaling for the first $\approx$~20\,s, although there are sizeable temporal oscillations observed. These may arise from oscillations within the building structure, although this has not been verified.

\end{document}